\documentclass[10pt, conference]{IEEEtran}
\usepackage{gensymb}
\usepackage{dsfont}
\usepackage{amsmath,amssymb,amsfonts}
\usepackage{epsfig}
\usepackage{cite}
\usepackage{hhline}
\usepackage{multirow}
\usepackage{framed}
\usepackage{xcolor}
\usepackage{makecell}
\usepackage{lipsum}
\usepackage{subcaption}
\usepackage{capt-of}
\usepackage[labelformat=simple]{subcaption}

\usepackage{enumerate}
\usepackage{enumitem}
\usepackage{color}
\usepackage{graphicx}
\usepackage{algorithmic}
\usepackage[ruled]{algorithm}
\usepackage{booktabs}
\usepackage{bm}
\usepackage{tikz}
\tikzset{%
    body/.style={inner sep=0pt,outer sep=0pt,shape=rectangle,draw=black},
    dimen/.style={<->,>=latex,thin,every rectangle node/.style={fill=white,midway,font=\sffamily}},
    symmetry/.style={dashed,thin},
}

\makeatletter
\newcommand{\vast}{\bBigg@{3}}
\newcommand{\Vast}{\bBigg@{4}}
\makeatother
\renewcommand{\IEEEQED}{\IEEEQEDopen}

\showoutput
\begin{document}

\title{On the Relay-Fallback Tradeoff in \\ Millimeter Wave Wireless System}

\author{\IEEEauthorblockN{Roberto Congiu\IEEEauthorrefmark{2}, Hossein Shokri-Ghadikolaei\IEEEauthorrefmark{3}, Carlo Fischione\IEEEauthorrefmark{3}, and Fortunato Santucci\IEEEauthorrefmark{2}}
\IEEEauthorblockA{\IEEEauthorrefmark{2}Centre of Excellence DEWS and DISIM, University of L'Aquila, L'Aquila, Italy\\
\IEEEauthorrefmark{3}Electrical Engineering, KTH Royal Institute of Technology, Stockholm, Sweden \\
{Emails: roberto.congiu@graduate.univaq.it, hshokri@kth.se, carlofi@kth.se, and fortunato.santucci@univaq.it}}}


\newtheorem{defin}{Definition}
\newtheorem{theorem}{Theorem}
\newtheorem{prop}{Proposition}
\newtheorem{lemma}{Lemma}
\newtheorem{corollary}{Corollary}
\newtheorem{alg}{Algorithm}
\newtheorem{remark}{Remark}
\newtheorem{result}{Result}
\newtheorem{example}{Example}
\newtheorem{notations}{Notations}
\newtheorem{assumption}{Assumption}

\newcommand{\be}{\begin{equation}}
\newcommand{\ee}{\end{equation}}
\newcommand{\ba}{\begin{array}}
\newcommand{\ea}{\end{array}}
\newcommand{\bea}{\begin{eqnarray}}
\newcommand{\eea}{\end{eqnarray}}
\newcommand{\combin}[2]{\ensuremath{ \left( \ba{c} #1 \\ #2 \ea \right) }}
\newcommand{\diag}{{\mbox{diag}}}
\newcommand{\rank}{{\mbox{rank}}}
\newcommand{\dom}{{\mbox{dom{\color{white!100!black}.}}}}
\newcommand{\range}{{\mbox{range{\color{white!100!black}.}}}}
\newcommand{\image}{{\mbox{image{\color{white!100!black}.}}}}
\newcommand{\herm}{^{\mbox{\scriptsize H}}}  
\newcommand{\sherm}{^{\mbox{\tiny H}}}       
\newcommand{\tran}{^{\mbox{\scriptsize T}}}  
\newcommand{\tranIn}{^{\mbox{-\scriptsize T}}}  
\newcommand{\card}{{\mbox{\textbf{card}}}}
\newcommand{\asign}{{\mbox{$\colon\hspace{-2mm}=\hspace{1mm}$}}}
\newcommand{\ssum}[1]{\mathop{ \textstyle{\sum}}_{#1}}

\newcommand{\vbar}{\raisebox{.17ex}{\rule{.04em}{1.35ex}}}
\newcommand{\vbarind}{\raisebox{.01ex}{\rule{.04em}{1.1ex}}}
\newcommand{\D}{\ifmmode {\rm I}\hspace{-.2em}{\rm D} \else ${\rm I}\hspace{-.2em}{\rm D}$ \fi}
\newcommand{\T}{\ifmmode {\rm I}\hspace{-.2em}{\rm T} \else ${\rm I}\hspace{-.2em}{\rm T}$ \fi}
\newcommand{\B}{\ifmmode {\rm I}\hspace{-.2em}{\rm B} \else \mbox{${\rm I}\hspace{-.2em}{\rm B}$} \fi}
\newcommand{\Hil}{\ifmmode {\rm I}\hspace{-.2em}{\rm H} \else \mbox{${\rm I}\hspace{-.2em}{\rm H}$} \fi}
\newcommand{\C}{\ifmmode \hspace{.2em}\vbar\hspace{-.31em}{\rm C} \else \mbox{$\hspace{.2em}\vbar\hspace{-.31em}{\rm C}$} \fi}
\newcommand{\Cind}{\ifmmode \hspace{.2em}\vbarind\hspace{-.25em}{\rm C} \else \mbox{$\hspace{.2em}\vbarind\hspace{-.25em}{\rm C}$} \fi}
\newcommand{\Q}{\ifmmode \hspace{.2em}\vbar\hspace{-.31em}{\rm Q} \else \mbox{$\hspace{.2em}\vbar\hspace{-.31em}{\rm Q}$} \fi}
\newcommand{\Z}{\ifmmode {\rm Z}\hspace{-.28em}{\rm Z} \else ${\rm Z}\hspace{-.38em}{\rm Z}$ \fi}

\newcommand{\sgn}{\mbox {sgn}}
\newcommand{\var}{\mbox {var}}
\newcommand{\E}{\mbox {E}}
\newcommand{\cov}{\mbox {cov}}
\renewcommand{\Re}{\mbox {Re}}
\renewcommand{\Im}{\mbox {Im}}
\newcommand{\cum}{\mbox {cum}}

\renewcommand{\vec}[1]{{\bf{#1}}}     
\newcommand{\vecsc}[1]{\mbox {\boldmath \scriptsize $#1$}}     
\newcommand{\itvec}[1]{\mbox {\boldmath $#1$}}
\newcommand{\itvecsc}[1]{\mbox {\boldmath $\scriptstyle #1$}}
\newcommand{\gvec}[1]{\mbox{\boldmath $#1$}}

\newcommand{\balpha}{\mbox {\boldmath $\alpha$}}
\newcommand{\bbeta}{\mbox {\boldmath $\beta$}}
\newcommand{\bgamma}{\mbox {\boldmath $\gamma$}}
\newcommand{\bdelta}{\mbox {\boldmath $\delta$}}
\newcommand{\bepsilon}{\mbox {\boldmath $\epsilon$}}
\newcommand{\bvarepsilon}{\mbox {\boldmath $\varepsilon$}}
\newcommand{\bzeta}{\mbox {\boldmath $\zeta$}}
\newcommand{\boldeta}{\mbox {\boldmath $\eta$}}
\newcommand{\btheta}{\mbox {\boldmath $\theta$}}
\newcommand{\bvartheta}{\mbox {\boldmath $\vartheta$}}
\newcommand{\biota}{\mbox {\boldmath $\iota$}}
\newcommand{\blambda}{\mbox {\boldmath $\lambda$}}
\newcommand{\bmu}{\mbox {\boldmath $\mu$}}
\newcommand{\bnu}{\mbox {\boldmath $\nu$}}
\newcommand{\bxi}{\mbox {\boldmath $\xi$}}
\newcommand{\bpi}{\mbox {\boldmath $\pi$}}
\newcommand{\bvarpi}{\mbox {\boldmath $\varpi$}}
\newcommand{\brho}{\mbox {\boldmath $\rho$}}
\newcommand{\bvarrho}{\mbox {\boldmath $\varrho$}}
\newcommand{\bsigma}{\mbox {\boldmath $\sigma$}}
\newcommand{\bvarsigma}{\mbox {\boldmath $\varsigma$}}
\newcommand{\btau}{\mbox {\boldmath $\tau$}}
\newcommand{\bupsilon}{\mbox {\boldmath $\upsilon$}}
\newcommand{\bphi}{\mbox {\boldmath $\phi$}}
\newcommand{\bvarphi}{\mbox {\boldmath $\varphi$}}
\newcommand{\bchi}{\mbox {\boldmath $\chi$}}
\newcommand{\bpsi}{\mbox {\boldmath $\psi$}}
\newcommand{\bomega}{\mbox {\boldmath $\omega$}}

\newcommand{\bolda}{\mbox {\boldmath $a$}}
\newcommand{\bb}{\mbox {\boldmath $b$}}
\newcommand{\bc}{\mbox {\boldmath $c$}}
\newcommand{\bd}{\mbox {\boldmath $d$}}
\newcommand{\bolde}{\mbox {\boldmath $e$}}
\newcommand{\boldf}{\mbox {\boldmath $f$}}
\newcommand{\bg}{\mbox {\boldmath $g$}}
\newcommand{\bh}{\mbox {\boldmath $h$}}
\newcommand{\bp}{\mbox {\boldmath $p$}}
\newcommand{\bq}{\mbox {\boldmath $q$}}
\newcommand{\br}{\mbox {\boldmath $r$}}
\newcommand{\bs}{\mbox {\boldmath $s$}}
\newcommand{\bt}{\mbox {\boldmath $t$}}
\newcommand{\bu}{\mbox {\boldmath $u$}}
\newcommand{\bv}{\mbox {\boldmath $v$}}
\newcommand{\bw}{\mbox {\boldmath $w$}}
\newcommand{\bx}{\mbox {\boldmath $x$}}
\newcommand{\by}{\mbox {\boldmath $y$}}
\newcommand{\bz}{\mbox {\boldmath $z$}}

\newenvironment{Ex}
{\begin{adjustwidth}{0.04\linewidth}{0cm}
\begingroup\small
\vspace{-1.0em}
\raisebox{-.2em}{\rule{\linewidth}{0.3pt}}
\begin{example}
}
{
\end{example}
\vspace{-5mm}
\rule{\linewidth}{0.3pt}
\endgroup
\end{adjustwidth}}


\maketitle

\begin{abstract}
Millimeter wave (mmWave) communications systems are promising candidate to support extremely high data rate services in future wireless networks. MmWave communications exhibit high penetration loss (blockage) and require directional transmissions to compensate for severe channel attenuations and for high noise powers. When blockage occurs, there are at least two simple prominent options: 1) switching to the conventional microwave frequencies (fallback option) and 2) using an alternative non-blocked path (relay option). However, currently it is not clear under which conditions and network parameters one option is better than the other. To investigate the performance of the two options, this paper proposes a novel blockage model that allows deriving maximum achievable throughput and delay performance of both options. A simple criterion to decide which option should be taken under which network condition is provided. By a comprehensive performance analysis, it is shown that the right option depends on the payload size, beam training overhead, and blockage probability. For a network with light traffic and low probability of blockage in the direct link, the fallback option is throughput- and delay-optimal. For a network with heavy traffic demands and semi-static topology (low beam-training overhead), the relay option is preferable.
\end{abstract}


\section{Introduction}
\label{Intro}
\IEEEPARstart{G}{rowing} demands for higher data rates in most of the existing  wireless applications, are leading to 3~Gbps peak data rate for cellular networks (LTE-Advanced~\cite{3gpp}) and 6~Gbps for short range networks (IEEE 802.11ac~\cite{fluke}). While these numbers are enough for many current applications, they are significantly smaller than the minimum requirements of future applications such as mobile backhauling/fronthauling and offloading (minimum 20~Gbps), 8K video transfer at smart homes (minimum 25~Gbps), and wireless backup links in data centers (minimum 40~Gbps)~\cite{licentiate}. To alleviate the imminent spectrum scarcity and to meet the growing demands for extremely high data rate applications, millimeter wave (mmWave)
communications appear as an essential future step of wireless communication. Current mmWave standards such as IEEE~802.11ad can support up to 6.7 Gbps and its extension, recently formed as IEEE 802.11ay,  is envisioned to have more than 20~Gbps data rate~\cite{licentiate}.

The channel attenuation in the mmWave bands is generally higher than that of the conventional microwave bands (below 6 GHz). Moreover, huge bandwidth of millimeter wave systems leads to orders of magnitude higher noise power accumulated at the receiver, which in turn can substantially  reduce signal-to-noise ratio (SNR). However, small wavelength of the mmWave signals  facilitates the integration of many antenna elements in the current size of radio chips, which promises a significant antenna gain both at the transmitter and at the receiver. These antenna gains can compensate both for the higher channel attenuation and for the higher noise power of mmWave systems, and provide the same (if not higher) SNR level as the traditional microwave networks. Having this SNR over huge bandwidth is the main result in supporting Gbps data rates with mmWave communications.

Despite the evident advantages mentioned above, deafness and blockage are two important problems in mmWave communications~\cite{mmwave}.
Deafness refers to a misalignment between the main antenna beams of a transmitter and receiver, prohibiting establishment of a high quality mmWave link. To avoid deafness, the transmitter and receiver should execute a time consuming beam-training procedure~\cite{5}.
Blockage is high penetration loss of the mmWave signals, e.g., 20-35~dB due to the human body and 20-30~dB due to the furniture in the environment~\cite{mmwave}.  These extreme attenuations cannot be efficiently compensated by increasing the transmission power. Upon appearance of an obstacle in the direct link between a transmitter and a receiver, there are two promising options to alleviate the blockage: relaying~\cite{relay}  and fallback~\cite{5}. In the \textit{relaying} technique, the source node bypasses the obstacles and communicates with the destination using an intermediate node, called relay. Therefore, the communication can be continued at the mmWave frequency. However, the source-relay and relay-destination pairs should undergo a beam-training procedure to avoid possible deafness. The other solution, called \textit{fallback}, is to switch to the microwave bands (e.g., 2.4 GHz)  during blockage and return to the mmWave bands once the obstacles disappears. In this case, there is no additional beam-training overhead at the expense of temporary  performance loss due to substantially smaller bandwidth of the microwave channel. To efficiently address blockage, the main question is whether a mmWave transmitter-receiver pair  should establish a new path using a relay node, which may cost extra beam-training overhead, or it should go for the fallback option, which may cost some temporary performance degradation.

To develop a proper decision criteria, we first need a blockage model that can capture the following four properties: (\textit{i}) randomness in the appearance and disappearance of the obstacles in one communication link, (\textit{ii}) having one obstacle on multiple angularly close communication links (angular correlation), (\textit{iii}) having multiple obstacles in a communication link, and (\textit{iv}) time correlation of the blockage events.
Using experimental mmWave channel measurements,~\cite{ref2} proposes a statistical blockage model that addresses (\textit{i}).
In~\cite{noiselimited}, a new blockage model for a general mmWave network is proposed that addresses properties (\textit{i}) and (\textit{ii}). Although all these models have been used for network throughput analysis, they lack of a thorough throughput and delay analysis for a mmWave network with dynamic environment. For instance, there is no distinction between having only one obstacle and having multiple obstacles in a link in~\cite{ref2,noiselimited}, (property (\textit{iii})). Moreover, an accurate delay analysis requires capturing the time correlation of blockage events in the blockage model, (property \textit(iv)). These drawbacks of existing literature are an obstacle for comparing different options in handling blockage.

In this paper, we investigate efficient solutions to address blockage. We first propose a new mathematical model for blockage. In particular, we model the wireless channel by a queue system, the appearance of an obstacle as a new message arrival, the disappearance of an obstacle by a message departure and therefore the blockage duration by the busy period of the server. Note that messages do not refer to real messages, but only to events of blockage occurrence: the term is misused to be consistent with the queuing network theory for communications. Using this model, we derive closed-form expressions, for the achievable throughput of both relay and fallback options with unbounded delay constraint (equivalently, achievable throughput for infinitely large messages).  We then develop a simple decision criteria to pick the best (throughput-optimal) option.  We show that it depends on the beam-training overhead, blockage duration and operating beamwidth. We then focus on finding the achievable throughput with finite delay requirements and highlight the existence of delay-throughput tradeoff, which may prohibit using the relay/fallback option for certain scenarios. Depending on the throughput and delay requirements, we characterize the conditions under which we can choose the relay option, the fallback option, or both. We numerically illustrate the proposed decision criteria for fixed throughput demands, showing which options fulfill the requirements with less delay.

The paper is organized as follows. In Sections~\ref{sec: system-model}, we introduce general assumptions and our novel blockage model. Section~\ref{sec: RelayOrFallBack} characterizes the throughput and delay performance, and thereby the best option to address blockage. Simulations results are presented in Section~\ref{sec: Numerical-results}, and the paper is concluded in Section~\ref{sec: Conclusion}.

\section{System Model}\label{sec: system-model}
\subsection{General Assumptions}
\begin{figure}[t]
\centering
\includegraphics[width=0.8\columnwidth]{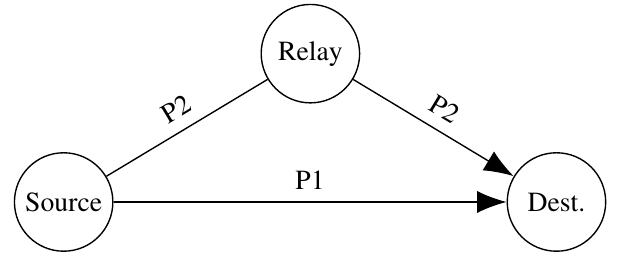}

\caption{The considered simple network: P1 is the direct path using either mmWave or micro wave channels, P2 is the relaying path using the mmWave channel.}
\label{sc}
\end{figure}
We consider a source node, a destination node, and a half-duplex relay, shown in Fig.~\ref{sc}. The communication link between the source and destination nodes can be established either by direct path, with millimeter and microwave frequency bands, or by the relay node in mmWave frequency bands. The best option to establish the communication link depends on the channel model, antenna gains, beam-training overhead of mmWave communications, payload size of the transmitted messages, and the quality-of-service requirements (in terms of throughput and delay).

The complete scenario would include several combination options: using microwave frequency bands also for the relay or using amplify and forward and decode and forward relaying schemes. Since this is the first study to analyze the issue, we are restricting ourselves to the simplest cases that are still of general and practical interest.
Without loss of generality, we consider a simple deterministic channel model that includes a constant attenuation at reference distance 1~meter and a distance dependent attenuation. Besides these attenuation sources, we consider a constant penetration loss $l_p$ due to any obstacle in the path. Note that we neglect the reflection for sake of simplicity, though we recently show that this assumption introduces a negligible error into the resulting interference model~\cite{Shokri2015WhatIs}. Let $d_{ij}$ be the distance between transmitter $i$ and receiver $j$ (transmitter can be either the source node or the relay node and the receiver can be either the relay node or the destination node), $c$ be the average attenuation at reference distance 1~meter, $n_j$ be the number of obstacles in this link, $l_o$ be the penetration loss due to any obstacle in dB, $\alpha$ be the attenuation factor due to atmospheric absorption. Then, the channel gain between transmitter $i$ and receiver $j$ is
\begin{equation}\label{eq: channel-gain}
g_{ij}^{\mathrm{x}} \:  [\mbox{dB}]= -c_{\mathrm{x}} - 20\log(d_{ij}) - 10 \log(e^{\alpha_{\mathrm{x}} d_{ij}}) - {n_j l_{o}}_{\mathrm{x}} \:,
\end{equation}
where subscript ${\mathrm{x}}$ is a label denoting microwave (${\mathrm{\mu}}$) or mmWave (${\mathrm{m}}$) band. Now, we present some values for the parameters of~\eqref{eq: channel-gain} in mmWave and in microwave. We have $\alpha_{\mathrm{\mu}}=$0.005~dB/Km at 2.4~GHz and $\alpha_{\mathrm{m}} =$16~dB/Km at 60~GHz~\cite{atmo}; about $c_{\mathrm{x}}$ we have 68~dB at 60~GHz and 46.7~dB at 2.4~GHz~\cite{agilent}.
We assume that the obstacles are impenetrable at the mmWave frequencies ($l_p \to \infty$). This means that, upon appearance of an obstacle on P1 in Fig.~\ref{sc}, source and destination nodes can reconfigure their antennas to use the relay node to pursue their mmWave communications or temporarily switch to the microwave communications. We assume that all nodes can instantly and perfectly detect appearance and disappearance of the obstacles on their active paths. For instance, the source and destination nodes detect appearance and disappearance of obstacles if P1 is active. Therefore, they can instantly react to any change in the underlying obstacle process. With this assumption, we can mathematically find the achievable throughput for unbounded and bounded delay cases.

For the beamforming model, we assume that all the nodes operate with the same beamwidth $\theta$. Without loss of generality, we consider an ideal sector antenna pattern with constant antenna gain $\varepsilon << 1$ inside the side lobe and another constant gain $2 \pi /\theta - \varepsilon ( 2 \pi - \theta )/\theta$ inside the main lobe~\cite{hosseinbeamforming}. Although the antenna gain in the main lobe substantially boosts the link budget, the transmitter and receiver should undergo a time consuming beam-training procedure to establish a sufficiently good mmWave link. In current mmWave standards, the transmitter and receiver start with finding the best sector-level beams (course solution) with a sequence of pilot transmissions, followed by finding the optimal beam-level beams (fine solution) within the selected sector. Clearly, achieving higher antenna gains requires adopting narrower beams, which in turn increases the beam-training overhead. As shown in~\cite{nitsche2015steering}, sector-level beam-training overhead can be well alleviated by tracking over time.
Let $\phi$ and $\theta$ be sector-level and beam-level beamwidths at the transmitter and at the receiver sides, respectively. Neglecting sector-level alignment overhead, and denoting the time required for a single pilot transmission by $T_p$, the beam-training overhead is~\cite{hosseinbeamforming}
\begin{equation}\label{align}
T_{a} (\theta)=\left \lceil{\frac{\phi}{\theta}}\right \rceil^2 T_p \:,
\end{equation}
where $\lceil{\cdot}\rceil$ is the ceiling function, returning the smallest following an integer.

Motivated by the high directionality of mmWave systems, we assume that the destination node is able to receive signal either from P1 or P2, not both. This is different from the conventional microwave networks, where the superposition of signals received from both paths is used for the decoding.
%

\begin{figure}[t]
\centering
\begin{tikzpicture}
    \node [body,minimum height=6mm,minimum width=2.8cm,anchor=south west] (body1) at (0,0) {\small{$X_0$}};
    \node [body,minimum height=6mm,minimum width=2.2cm,anchor=south west] (body2) at (2.8,0) {\small{$Y_0$}};
    \node [body,minimum height=6mm,minimum width=1.5cm,anchor=south west] (body3) at (5,0) {\small{$X_1$}};
    \node [body,minimum height=6mm,minimum width=2.1cm,anchor=south west] (body4) at (6.5,0) {\small{$Y_1$}};
    \draw (body1.south west) -- ++(0,-0.5) coordinate (D1) -- +(0,-2pt);
    \draw (body1.south east) -- ++(0,-0.5) coordinate (D2) -- +(0,-2pt);
    \draw [dimen] (D1) -- (D2) node {\footnotesize{$\mathrm{non}$-$\mathrm{LoS~period}$}};
    \draw (body2.south west) -- ++(0,-0.5) coordinate (D3) -- +(0,-2pt);
    \draw (body2.south east) -- ++(0,-0.5) coordinate (D4) -- +(0,-2pt);
    \draw [dimen] (D3) -- (D4) node {\footnotesize{$\mathrm{LoS~period}$}};
    \draw (body1.south west) -- ++(0,+1) coordinate (D5) -- +(0,+5pt);
    \draw (body2.south east) -- ++(0,+1) coordinate (D6) -- +(0,+5pt);
    \draw [dimen] (D5) -- (D6) node {\footnotesize{$T_0$}};
    \draw (body3.south west) -- ++(0,+1) coordinate (D7) -- +(0,+5pt);
    \draw (body4.south east) -- ++(0,+1) coordinate (D8) -- +(0,+5pt);
    \draw [dimen] (D7) -- (D8) node {\footnotesize{$T_1$}};
\end{tikzpicture}

\caption{Time slots divided in phases. $T_i$ is $i$-th inter LoS interval.}
\label{tsl}
\end{figure}
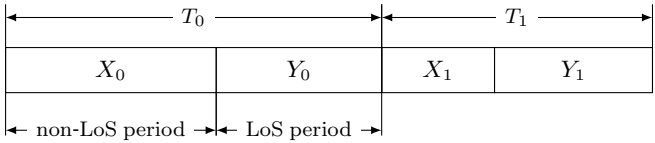

\subsection{Blockage Model}
To develop a systematic blockage model that meets all four properties of a proper blockage model, mentioned in Section~\ref{Intro}, we borrow some concepts from queuing theory~\cite{queue}. In particular, we model blockage of any link by a queue system, hereafter called \emph{blockage queue}, wherein a new message arrival means the appearance of a new obstacle in the link and a new message departure means the disappearance of an obstacle. Let $x(t)$ denote the number of obstacles in the blockage queue at time $t$. If $x(t)=0$, we say this link is in the line-of-sight (LoS) condition, if $x(t)\neq 0$ it is in the non-LoS condition. Consider a reference time. We define $i$-th \textit{virtual time slot}, denoted by $T_i$, as the time interval between the beginning of $i$-th and $(i+1)$-th non-LoS conditions, namely $T_i = t_{i+1} - t_{i}$ such that $x_{t_{i}-\epsilon}=0$ for infinitesimal $\epsilon>0$ but $x_{t_{i}}>0$, and that $x_{t_{i+1}-\epsilon}=0$ but $x_{t_{i+1}}>0$. We define $X_i$ as $i$-th non-LoS period during which $x(t)>0$, and similarly define $Y_i$ as $i$-th LoS period during which $x(t)=0$. In other words, $X_i$ and $Y_i$ respectively show $i$-th non-LoS and LoS periods, and $T_i = X_i + Y_i$. These periods are illustrated in Fig.~\ref{tsl}. Depending on the arrival and departure processes of the obstacles (their numbers, sizes, and mobilities), $X_i$ and $Y_i$ are random variables, whose distributions depend on the \textit{busy period} (non-LoS) and \textit{idle period} (LoS) of the blockage queue system. We assume that $X_i$ and $Y_i$ individually are i.i.d. random variables, thus with a slight abuse of notation we remove the index $i$ for the rest of the paper. The random variables $X$ and $Y$ give the random variable $T=X+Y$.

We consider the special class of Poisson arrival and Poisson departure of the obstacles; however, the analysis can be easily extended to any other arrival and departure distributions. Let $\lambda$ and $\nu$ be the arrival and departure rates. We note that every arrived obstacle should experience immediate service and should neither wait nor depend on other obstacles currently being served in the blockage queue system. Therefore, we consider an $M|M|\infty$ blockage queue system. To evaluate $X$ and $Y$, we use the notion of \textit{congestion period} of $M|M|\infty$ systems. A $c$-congestion period is the interval starting when an arriving customer (obstacle in our scenario) finds $c$ customers in the system, until a departing customer leaves $c$ customers behind. Clearly, $X$ corresponds to $0$-congestion period. Thus~\cite[Equation (3.1)]{queue},
\begin{equation}\label{xmean2}
\mathbb{E}[X]=\frac{1}{\lambda}\left(e^{\frac{\lambda}{\nu}}-1\right) \:,
\end{equation}
where $\mathbb{E}$ is the expectation operation. Furthermore, for Poisson obstacle arrivals,
\begin{equation}\label{ymean}
\mathbb{E}[Y]=\frac{1}{\lambda}\:.
\end{equation}
We assume that there is no obstacle on the relay path for mathematical tractability, though the analysis of this paper can be readily extended for the general case of having blockage on all mmWave links. For the rest of the paper, $i$-th virtual time slot is defined based on arrival and departure of the obstacles on the direct link between the source and destination nodes.
In the following section, we use this blockage model to analyze the throughput and delay of our mmWave network.

\section{Relay or Fallback}\label{sec: RelayOrFallBack}
\subsection{Throughput Analysis for Long Packets Scenario}
To analyze the achievable throughput of both relaying and fallback options, we define the following transmission rates. We denote by $C_{\mathrm{d \mu}  X}$ the achievable transmission rate at the direct path P1 at microwave band during $X$, by $C_{\mathrm{d m} X}$ the achievable transmission rate at the mmWave band during $X$, and by $C_{\mathrm{r m}}$ the achievable transmission rate at the relay path P2 at the mmWave band during $T = X + Y$. In the fallback option, the source transmits with rate $C_{\mathrm{d \mu} X}$ during $X$ and with rate $C_{\mathrm{d m} Y}$ during $Y$. This procedure will be repeated in all virtual time slots. It follows that the throughput of the fallback option at $i$-th virtual time slot is
$R_{i}^{\text{fallback}} = C_{\mathrm{d \mu}  X_i}X_i+C_{\mathrm{d m}  Y_i}Y_i$ and thus
\begin{equation}\label{eq: fallback-average-throughput}
\mathbb{E} \left[R^{\text{fallback}}\right] = C_{\mathrm{d \mu}  X} \, \mathbb{E}[X] \, + \,  C_{\mathrm{d m} Y} \, \mathbb{E}[Y] \:.
\end{equation}
where $\mathbb{E}[X] $ and $\mathbb{E}[Y] $ are given in~\eqref{xmean2} and~\eqref{ymean}, respectively. To derive $C_{\mathrm{d \mu}  X}$ and $C_{\mathrm{d \mu}  Y}$, we denote by $\mathrm{SNR}_{\mathrm{d \mu}}$ and $\mathrm{SNR}_{\mathrm{d m}}$ the signal-to-noise ratio (SNR) of the direct path at the microwave and mmWave bands, and by $W_{\mathrm{\mu}}$ and $W_{\mathrm{m}}$ the bandwidth of the microwave and mmWave bands, respectively. Then, the maximum achievable rates for sufficiently large packets (capacity of the channel) are equal to $C_{\mathrm{d \mu}  X} = {W_{\mathrm{\mu}}} \, \mathrm{\log_2(1+\mathrm{SNR}_{\mathrm{d \mu}})}$ bits per second during $X$ and $C_{\mathrm{d m}  Y} = {W_{\mathrm{m}}} \, \mathrm{\log_2(1+\mathrm{SNR}_{\mathrm{d m}})}$ bits per second during $Y$. Let $n$ be the power of white Gaussian noise and $p$ be the transmission power of the transmitter. Then, assuming omnidirectional transmission in microwave (to be robust to blockage) and directional transmission in mmWave, we have $\mathrm{SNR}_{\mathrm{d \mu}} = p {g_{j,k}^{\mathrm{\mu}}}/{n}$ and
\begin{equation}\label{eq: SNR-mmWave}
\mathrm{SNR}_{\mathrm{d m}} = \frac{p}{n} g_{j,k}^{\mathrm{m}} \left(\frac{2 \pi - \varepsilon \left( 2 \pi - \theta \right)}{\theta}\right) \:,
\end{equation}
where $j$ and $k$ show the source and destination nodes, and ${g_{j,k}^{\mathrm{\mu}}}$ and ${g_{j,k}^{\mathrm{m}}}$ are given by~\eqref{eq: channel-gain}. Now, we formulate $C_{\mathrm{r m}}$.

In the relay option, the source-relay-destination path will be firstly established at the mmWave band, which takes $T_{a}$. Then, the source node transmits at rate $C_{\mathrm{r m}}$ toward the relay node for half of the remaining $T - T_a$ and the remaining time is used for the relay-destination link. To evaluate the data rate between the source and destination nodes in a multihop communication system, achievable rates of each hop [in bps] should be multiplied by the corresponding available transmission time. Then, the number of transmitted bits between source and destination is the minimum of the throughput of all hops. Dividing this number of bits by the total slot duration gives the overall rate [in bps]. In our case, since both hops have the same transmission time, we have
\begin{equation}\label{c2}
C_{\mathrm{rm}} = {W_{\mathrm{m}}} \times \min\Big(\log_2\left(1+\mathrm{SNR}_{jk}\right) ,  \log_2\left(1+\mathrm{SNR}_{kl}\right)\Big)\:,
\end{equation}
where $\mathrm{SNR}_{jk}$ is the SNR of source-relay link, and $\mathrm{SNR}_{kl}$ is the SNR of the relay-destination link. These SNR values can be found by substituting the corresponding link length into~\eqref{eq: SNR-mmWave}. Finally, we can find the throughput of the relay option in $i$-th virtual time slot, given no obstacle on the relay path, is
\begin{equation}\label{eq: throughput-relay}
R^{\text{relay}} = C_{\mathrm{rm}}  \times \max\left(\frac{X + Y - T_a}{2},0\right) \:.
\end{equation}
This equation implies that the beam-training time should not exceed the total duration of a virtual time slot, otherwise the throughput of that slot would be 0. Then,
\begin{align}\label{eq: throughput-relay2}
\hspace{-1mm} \mathbb{E} \left[\hspace{-0.7mm}R^{\text{relay}}\right] \hspace{-1mm} = \hspace{-0.7mm} C_{\mathrm{rm}} \hspace{-0.7mm} \int_{x=0}^{\infty}\int_{y=\max(T_a-x,0)}^{\infty} \hspace{-2mm}\frac{x+y-T_a}{2} f_{XY}(x,y)\,\mathrm{d}x \mathrm{d}y
\end{align}
where $f_{XY}(x,y)$ is joint distribution of the obstacle arrival and departure processes. In general, $f_{XY}(x,y)$ is unknown for an $M|M|\infty$ system, prohibiting mathematical calculation of~\eqref{eq: throughput-relay2}. Nonetheless, we note that random variables $X$ and $Y$ model the presence duration of the obstacles, so their typical values would be, at least, a few seconds. However, $T_a$ is the alignment overhead, which ranges around few hundreds of micro seconds. This is orders of magnitude smaller than $X$ and $Y$. Therefore, we take the following assumptions, which greatly simplify the analysis with negligible accuracy loss. We comment on the accuracy of this assumption in Section~\ref{sec: Numerical-results}.

\begin{assumption}\label{assumption: XYT_a}
We assume that $X+Y>T_a$.
\end{assumption}
With Assumption~\ref{assumption: XYT_a} and from~\eqref{eq: throughput-relay}, we have
\begin{equation}\label{eq: ExpectedThroughputRelay}
\mathbb{E} \left[R^{\text{relay}}\right] = C_{\mathrm{rm}} \, \frac{\mathbb{E} \left[X + Y\right]  - T_a}{2}\,.
\end{equation}
Later in Section~\ref{sec: Numerical-results}, we validate Assumption~\ref{assumption: XYT_a} by observing orders of magnitude difference between $\min(T)$ and $T_a$, even for extremely narrow beams (e.g., $\theta=1\degree$), which corresponds to the very high $T_a$.

\subsection{Throughput Optimal Criteria to Address Blockage}
The throughput-optimal decision rule adopts the option that offers higher average throughput, namely
\begin{equation*}
\mathbb{E} \left[R^{\text{fallback}}\right] \underset{\text{Relay}}{\overset{\text{Fallback}}{\gtrless}} \mathbb{E} \left[R^{\text{relay}}\right] \:,
\end{equation*}
which, by substituting~\eqref{eq: fallback-average-throughput} and~\eqref{eq: ExpectedThroughputRelay}, leads to the following throughput-optimal decision rule:
\begin{equation}\label{thro3}
\frac{C_{\mathrm{rm}}}{2}T_a \underset{\text{Relay}}{\overset{\text{Fallback}}{\gtrless}}
\left(\frac{C_{\mathrm{rm}}}{2} - C_{\mathrm{d \mu}  X}\right) \mathbb{E}[X] + \left( \frac{C_{\mathrm{rm}}}{2}-C_{\mathrm{d m} Y}\right) \mathbb{E}[Y] \:,
\end{equation}
where $T_a$, $\mathbb{E}[X]$, and $\mathbb{E}[Y]$ are given in~\eqref{align}--\eqref{ymean}.

So far, we analyzed an optimal decision rule for a simplistic system with sufficiently large packets (which allowed to use the Shannon capacity equation) and with infinitely large backlogged packets. In the following, we consider a more realistic wireless system with short packets, which is usually the case in machine-type communications, and random packet arrival with an exponential distribution.

\subsection{Throughput and Delay Analysis for Short Packets Scenario}
As shown in~\cite{coderate}, with short packets we may not be able to achieve the Shannon capacity, which affects the transmission rates in both options. Let $P_b$ be the packet error probability and $L$ be the packet size. Then, the achievable rate instead of being $W \log_2(1+\mathrm{SNR})$, with corresponding bandwidth and SNR, would be~\cite{coderate}
\begin{equation}\label{blocklength}
W \left(\log_2(1+\mathrm{SNR}) - \sqrt{\frac{V}{L}}Q^{-1}(P_b) + \mathrm{\log_2}(L) \right) \:,
\end{equation}
where
\begin{align}
V = \frac{\mathrm{SNR}}{2}\frac{\mathrm{SNR}+2}{\left(\mathrm{SNR}+1\right)^2} \, \log_2 (e)\:.
\label{blocklength2}
\end{align}
We apply this modification to derive the right $C_{\mathrm{rm}}$, $C_{\mathrm{d \mu}  X}$, and $C_{\mathrm{d \mu}  Y}$ in the short packet scenario.

To capture the tradeoff between throughput and delay, and without loss of generality, we model a transmitter-receiver pair by an $M|D|1$ queue systems.\footnote{This queue system, which models the data transmission, is independent of the blockage queue model.} Specifically, the bits generated and transmitted by the source follow the Poisson distribution with rate $G$. The source has an infinite buffer-length, so all the bits are eventually served. The constant service times $D_{\text{fallback}}$ for the fallback option and $D_{\text{relay}}$ for the relay option are~\cite{queue},
\begin{equation}\label{Delay11}
D_{\text{fallback}} = \left[\left(C_{\mathrm{d \mu}  X}\frac{\mathbb{E}[X]}{\mathbb{E}[T]} + C_{\mathrm{d m}  Y}\frac{\mathbb{E}[Y]}{\mathbb{E}[T]}\right)\right]^{-1} \:,
\end{equation}
\begin{equation}
D_{\text{relay}} = \left[C_{\mathrm{rm}} \, \frac{\mathbb{E}[X+Y]}{2\mathbb{E}[T]}\right]^{-1} \:.
\end{equation}
Then, the average delay by two options are given by
\begin{equation}\label{eq: delay}
\tau_{\mathrm{x}} = \frac{1}{2S_{\mathrm{x}}}\cdot \frac{2-\rho_{\mathrm{x}}}{1-\rho_{\mathrm{x}}}\:,
\end{equation}
where ${\mathrm{x}}$ is a label denoting either the fallback or the relay option, $\rho_{\mathrm{x}}={G_{\mathrm{x}}}/{S_{\mathrm{x}}}$ is the utilization, $G_{\mathrm{x}}$ is the bit arrival rate, and $S_{\mathrm{x}}={1}/{D_{\mathrm{x}}}$ is the average bit departure rate.
As long as the queue system remains stable, namely $\rho_{\mathrm{x}}<1$, the system can serve all the incoming data served with delay $\tau_{\mathrm{x}}$. Therefore, the throughput of such stable system is $G_{\mathrm{x}}$.
Once the system becomes unstable, e.g., due to very high arrival rate, the throughput is bounded by $S_{\mathrm{x}}$, while the delay grows large.
\subsection{Throughput-delay Optimal Criteria to Address Blockage}
Considering the throughput and corresponding delay, formulated in~\eqref{blocklength}--~\eqref{eq: delay}, we can find the optimal option to address blockage. For a given delay requirements, the option that provides higher throughput is preferable. Alternatively, for a given throughput requirements, the option with lower delay is preferable. We numerically show the optimal decision rule in the next section.

\section{Numerical Results}\label{sec: Numerical-results}
To numerically illustrate the optimal option to alleviate blockage, we simulate a short range indoor wireless network, composed of three nodes working at 60~GHz. We consider 2.5~mW maximum transmission power, 90$\degree$ sector-level beams, and $\varepsilon=0.05$ antenna side lobe gain. The source-destination distance is 10~m and the relay node is located 10~m from both the source and destination nodes.
We assume $W_{\mathrm{\mu}} = 20$~MHz and $W_{\mathrm{m}} = 2.16$~GHz, so the noise powers are -101~dBm for the microwave channel and -80.7~dBm for the mmWave channel.
We consider $T_p = 20~\mu $s single pilot transmission time~\cite{802_15_3c}.
We define the \emph{normalized blockage time} as the time interval over which the direct channel is in the non-LoS condition, namely $\mathbb{E}[X/T]$. We assume $\lambda = 0.5$ obstacles/s arrival rate and sweep and the departure rate from 0.5 to 1 obstacles/s. This leads to variations of $\mathbb{E}[X]$, and consequently variations of the average blockage time. With the relay option, the source-destination connection is active at the mmWave band for less than half of the time slot; whereas with the fallback option, it is active at the microwave band during $X$ and at the mmWave band during $Y$.
The curves obtained by~\eqref{eq: throughput-relay2} and by~\eqref{eq: ExpectedThroughputRelay} are essentially identical because $T_a$ is orders of magnitude smaller than $T$, thus we only report the curves obtained by~\eqref{eq: ExpectedThroughputRelay}. The exact values of $T_a$ and $\min(T)$ are reported for each figure.

\begin{figure}[t]
\centering
\includegraphics[width=\columnwidth]{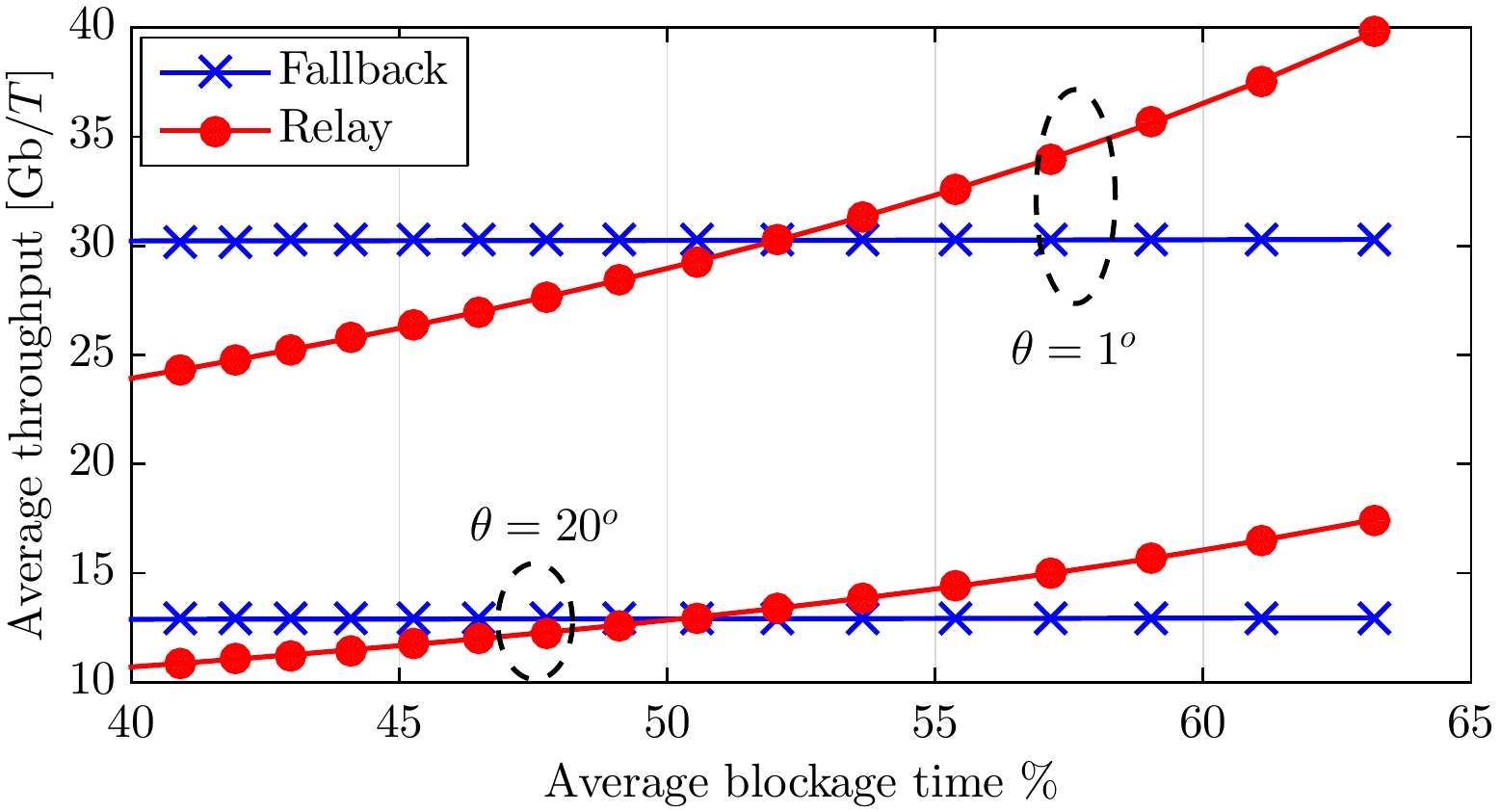}
\caption{The average throughput of the fallback and relay options per inter-LoS interval $T$, as computed by~\eqref{eq: fallback-average-throughput} and~\eqref{eq: throughput-relay2}, $\min(T)=3.4$~s, $T_{a}=0.5$~ms for $\theta = 20\degree$ and $T_{a}=0.16$~s for $\theta = 1\degree$.}
\label{th1}
\end{figure}
Assuming long packets and full buffers, Fig.~\ref{th1} shows the average throughput per inter-LoS interval $T$ against the normalized blockage time. With long packets and narrow beamforming, the mmWave transmission rate $C_{\mathrm{d m}  Y}$ is much higher than that of the microwave $C_{\mathrm{d \mu}  X}$. Therefore, throughput in the LoS period $Y$ dominates the throughput of the fallback option. As changing the average blockage time does not affect $\mathbb{E}[Y]$, see~\eqref{ymean}, the throughput of the fallback option is almost independent of the average blockage time. Higher $X$, however, increases $T$ and therefore increases the throughput of the relay option, see~\eqref{eq: throughput-relay2} and~\eqref{eq: ExpectedThroughputRelay}. With operating beamwidth $\theta=20\degree$, the alignment overhead is almost negligible compared to $T$. Therefore, the relay option can transmit at the mmWave band for $T/2$. Now, if the average blockage in the direct path is below 50\%, it is beneficial to use the fallback option; otherwise, the relay option is throughput-optimal. With narrower beamwidths, e.g., $\theta=1\degree$, the alignment overhead reduces the transmission time of the relay option. Consequently, the fallback option remains throughput-optimal for higher average blockage time values, as shown in Fig.~\ref{th1}. Moreover, notice that narrower beamwidth provides higher antenna gains, leading to higher average throughput for both options. Finally, Assumption~\ref{assumption: XYT_a} holds due to significant difference between $\min(T)$ and $T_a$ for all beamwidths.

\begin{figure}[t]
\centering
\includegraphics[width=\columnwidth]{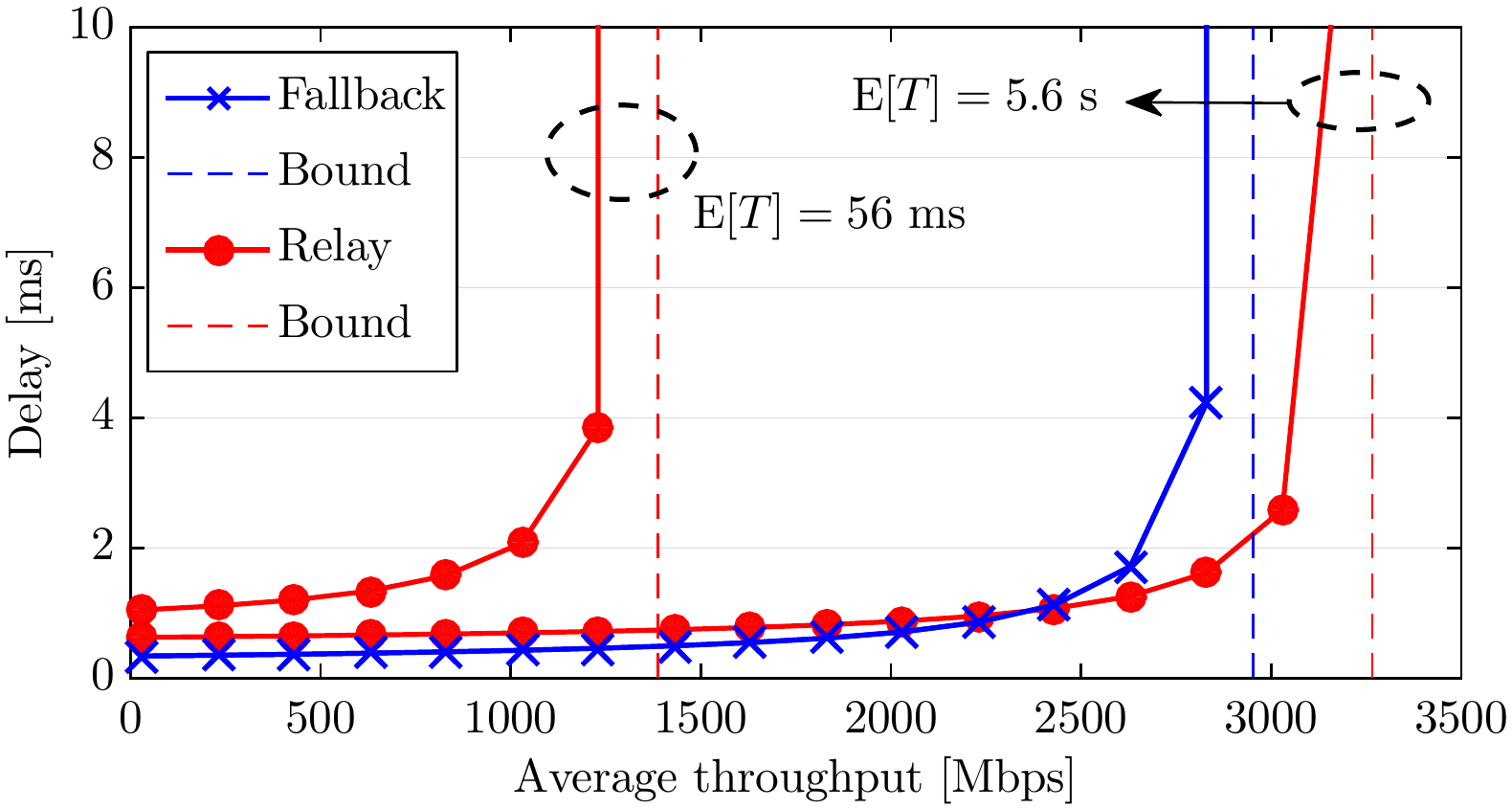}

\caption{Illustration of the delay-Throughput tradeoff for $\theta = 20 \degree$ and blockage time = 55\%. $\mathbb{E}[T]$ is reported in the figures. Upper bounds of the throughput in the fallback and relay options are shown by dashed lines.}
\label{delay1}
\end{figure}

Assuming the short packets scenario, Fig.~\ref{delay1} illustrates the delay-throughput tradeoff for $\theta = 20\degree$ and 55\% average blockage time. With $\mathbb{E}[T]=5.6$~s, the fallback option outperforms the relay option until 2250~Mbps data rate. Soon after this point, the fallback option cannot stably support the incoming traffic, leading to an exponential delay penalty. With shorter inter-LoS intervals, frequent executions of the alignment procedure makes the relay option less attractive. In the case of $\mathbb{E}[T]=56$~ms, the relay option becomes unstably serve the input traffic above 1.3~Gbps; whereas the falls back option supports until 2.8~Gbps data rate with bounded delay. Notice that this rate cannot be supported (with any delay) by the relay option in this example. In general, for light traffic, the fallback option is generally delay-optimal; whereas for heavy traffic applications the relay option may be delay-optimal if it does not require frequent re-alignments.

\begin{figure}[t]
\centering
\includegraphics[width=\columnwidth]{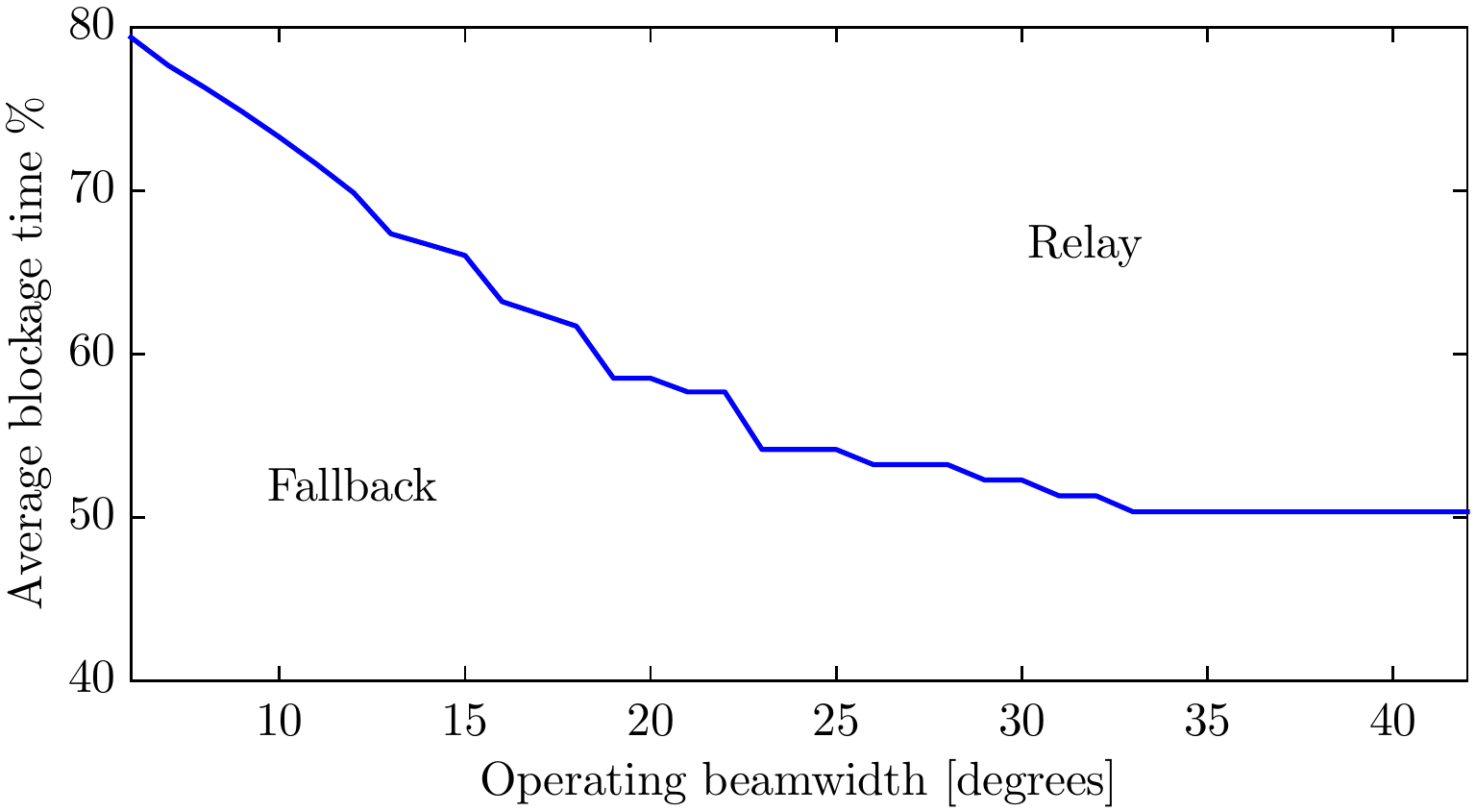}

\caption{The delay-optimal decision region to support 2~Gbps data rate.}
\label{high_payload}
\end{figure}
Fig.~\ref{high_payload} shows the decision region over which the fallback or relay option provides lower delay for 2~Gbps data rate. Note that either the fallback or relay option may not support this rate for some combinations of the blockage time and operating beamwidth, as illustrated in Fig.~\ref{delay1}. For narrow beams, the relay option is preferable only if the the direct link is blocked most of the time. For instance, with $\theta=1 \degree$, the average blockage should be above 80\% of the entire transmission time to choose the relay option. Wider operating beamwidths alleviate the alignment overhead, and therefore the relay option may be preferred for lower blockage time.

\section{Conclusion}\label{sec: Conclusion}
To support Gbps data rate without service disruption, millimeter wave (mmWave) systems must address blockage (high penetration loss). To this end, we first proposed a novel blockage model, based on which we investigated throughput and delay performance of two promising solution approaches: \emph{i}) switching to the conventional microwave frequencies during the non-line-of-sight period (\emph{fallback}) and \emph{ii}) using an alternative non-blocked path (\emph{relay}). We illustrated simple decision criteria to find the throughput-and delay-optimal approach to address blockage under various traffic patterns, beam-training overheads, and blockage probabilities. For high mobility scenarios with frequent re-executions of the beam-training procedure, the fallback option is throughput- and delay-optimal. This option is also preferred under light traffic conditions and low probability of blockage in the direct link. The relay option is better once the direct link is subject to high blockage probability or the incoming traffic is so heavy that the transmitter should not switch to the low capacity microwave link.
The results of this paper provide useful insights for the standardization of future mmWave systems.

For future work, we aim to investigate the throughput and delay performance of mitigating blockage by first-order reflection~\cite{relay}, and extend the derived decision rules.

\bibliographystyle{IEEEtran}
\bibliography{References}

\end{document}